\documentclass[twocolumn,showpacs,preprintnumbers,amsmath,amssymb]{revtex4}

\usepackage{graphicx}
\usepackage{dcolumn}
\usepackage{bm}

\begin{document}

\preprint{}

\title{Coherence of the Borromean three-body F\"orster resonances in Rydberg atoms}

\author{I.~I.~Ryabtsev$^{1, 2}$}
  \email{ryabtsev@isp.nsc.ru}
\author{I.~I.~Beterov$^{1, 2, 3}$}
\author{D.~B.~Tretyakov$^{1, 2}$}
\author{E.~A.~Yakshina$^{1, 2}$}
\author{V.~M.~Entin$^{1, 2}$}
\author{P.~Cheinet$^{4}$}
\author{P.~Pillet$^{4}$}

\affiliation{$^1$Rzhanov Institute of Semiconductor Physics SB RAS, 630090 Novosibirsk, Russia }
\affiliation{$^2$Novosibirsk State University, 630090 Novosibirsk, Russia}
\affiliation{$^3$Novosibirsk State Technical University, 630073 Novosibirsk, Russia}
\affiliation{$^4$Laboratoire Aim\'e Cotton, CNRS, Univ. Paris-Sud, ENS-Cachan, Universit\'e Paris-Saclay, 91405 Orsay, France}

\date{23 July 2018}

\begin{abstract}

We have observed recently the Stark-tuned three-body F\"orster resonances ${\rm 3}\times nP_{3/2} (|M|)\to nS_{1/2} +(n+1)S_{1/2} +nP_{3/2} (|M^{*} |)$ at long-range interactions of a few cold Rb Rydberg atoms [D.~B.~Tretyakov et al., Phys. Rev. Lett. \textbf{119}, 173402 (2017)]. The three-body resonance appears at a different dc electric field with respect to the ordinary two-body resonance ${\rm 2}\times nP_{3/2} (|M|)\to nS_{1/2} +(n+1)S_{1/2} $ and corresponds to a transition when the three interacting atoms change their states simultaneously (two atoms go to the \textit{S} states, and the third atom remains in the \textit{P} state but changes its moment projection), with the negligible contribution of the two-body resonance to the population transfer. It thus has a Borromean character and represents an effective three-body operator, which can be used to directly control the three-body interactions in quantum simulations and quantum gates implemented with Rydberg atoms. In this paper we theoretically investigate the coherence of such three-body resonances and we show that high-contrast Rabi-like population oscillations are possible for the localized Rydberg atoms in a certain spatial configuration. This paves the way to implementing three-qubit quantum gates and quantum simulations based on three-body Rydberg interactions.

\end{abstract}

 \maketitle

\section{Introduction}

Highly excited Rydberg atoms [1] are attractive for the development of quantum computers and simulators due to their strong long-range interactions [2-13]. The interactions are typically described by a two-body operator of dipole-dipole interaction for each pair of atoms in the atom ensemble [1]. Some quantum gates and simulations, however, demand to simultaneously control the interactions of three atoms [14-19]. This demands a three-body quantum operator that changes the states of the three atoms simultaneously and makes them all entangled. 

Such an operator has been implemented recently by us as a Borromean three-body F\"orster resonance ${\rm 3}\times nP_{3/2} (|M|)\to nS_{1/2} +(n+1)S_{1/2} +nP_{3/2} (|M^{*} |)$ for \textit{N}=3-5 cold Rb Rydberg atoms with with the principal quantum number \textit{n}=36, 37 [20]. We have found clear evidence that there is no signature of the three-body F\"orster resonances for exactly two interacting Rydberg atoms, while it is present for the larger number of atoms. 

The three-body resonances were first observed and explained in Ref.[21] for an ensemble of $\sim 10^5$ cold Cs Rydberg atoms. A three-body resonance corresponds to a transition when the three interacting atoms change their states simultaneously (two atoms go to the neighboring \textit{S} states, and the third atom remains in the \textit{P} state but changes its moment projection). In these resonances, one of the atoms carries away an energy excess preventing the two-body resonance, leading thus to a Borromean type of F\"orster energy transfer. The Borromean character means that the ordinary two-body resonance ${\rm 2}\times nP_{3/2} (|M|)\to nS_{1/2} +(n+1)S_{1/2} $ gives a negligible contribution to the population transfer, as the three-body resonance appears at a different dc electric field with respect to the two-body resonance. It thus represents an effective three-body operator, which can be used to directly control the three-body interactions in quantum simulations and quantum gates with Rydberg atoms.

In this paper we theoretically investigate the coherence of the Borromean three-body F\"orster resonances. This issue rises since in our experiment [20] the three-body resonances were rather broad and partially overlapped with the two-body resonances when the controlling dc electric field was scanned [see Fig.~1(b)], because the three atoms were randomly placed in a single interaction volume and their interaction energy was not fixed. The observed broadening and overlapping make it unclear if quantum gates and simulations are really possible with our three-body resonances, as the coherence should be conserved during the gate or simulation time. In particular, one needs to study if the Rabi-like population oscillations are possible for well localized Rydberg atoms in various spatial configurations, when the interaction energy is well fixed. If we find that such oscillations are possible, they will pave the way to developing the schemes of the three-qubit quantum gates (e.g., the Toffoli or Fredkin quantum gates [22,23]) and of the quantum simulators based on three-body Rydberg interactions [14-19].

\section{Experimental observations for disordered atoms in a single interaction volume}

Our experiment in Ref.~[20] was performed with cold $^{85}$Rb atoms in a magneto-optical trap [4]. It featured atom-number-resolved measurement of the signals obtained from \textit{N}=1-5 Rydberg atoms detected by a selective field ionization with a detection efficiency of \textit{T}$\approx$70\%. The normalized \textit{N}-atom signals $S_N$ were the fractions of atoms that have undergone a transition to the final \textit{nS} state. The recorded F\"orster resonance spectra were additionally processed to extract the true multiatom spectra $\rho_i$ (\textit{i}=1-5) taking into account finite detection efficiency [24]. The excitation of Rb atoms to the $nP_{3/2}$ Rydberg states was realized via the three-photon transition $5S_{1/2} \to 5P_{3/2} \to 6S_{1/2} \to nP_{3/2} $  by means of three cw lasers modulated to form 2 $\mu $s exciting pulses at a repetition rate of 5~kHz. A small Rydberg excitation volume of $\sim$15-20 $\mu$m in size was formed using the crossed tightly-focused laser beams. 

Figure~1(a) presents the numerically calculated Stark structure of the F\"orster resonance ${\rm 3}\times 37P_{3/2} \to 37S_{1/2} +38S_{1/2} +37P_{3/2}^{*} $ for three Rb Rydberg atoms. The energies \textit{W} of various three-body collective states are shown versus the controlling dc electric field. The intersections between collective states (labeled by numbers) correspond to the F\"orster resonances of various kinds. Intersections 2-7 are, in fact, two-body resonances that do not require the third atom and can be observed for two or more atoms. In such resonances, the dipole-dipole interaction induces transitions from the initial $37P_{3/2}$ state to the final $37S_{1/2}$ and $38S_{1/2}$ states in two of the three atoms, while the third atom remains in its initial \textit{P} state that does not change. Intersections 1 and 8 are three-body resonances occurring only in the presence of the third atom that carries away an energy excess preventing the two-body resonance [21]. The three-body resonances are distinguished from the two-body ones by the fact that the third atom does not remain in its initial \textit{P} state as its initial moment projection ($|M|=1/2$ or $|M|=3/2$) changes to the other one ($|M^*|=3/2$ or $|M^*|=1/2$, correspondingly). Therefore, the three-body resonance corresponds to the transition when the three interacting atoms change their states simultaneously. 

\begin{figure}
\includegraphics[scale=0.7]{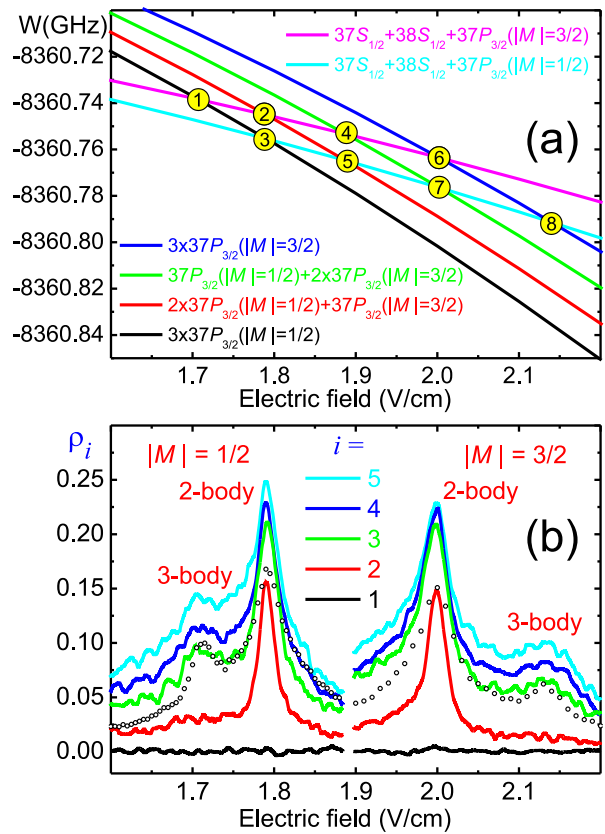}
\caption{\label{Fig1} (a) Numerically calculated Stark structure of the F\"orster resonance ${\rm 3}\times 37P_{3/2} \to 37S_{1/2} +38S_{1/2} +37P_{3/2}^{*} $ for three Rb Rydberg atoms. The energies \textit{W} of various three-body collective states are shown versus the controlling electric field. Intersections between collective states (labeled by numbers) correspond to the F\"orster resonances of various kinds. Intersections 2-7 are, in fact, two-body resonances that do not require the third atom. Intersections 1 and 8 are three-body resonances occurring only in the presence of the third atom that carries away an energy excess preventing the two-body resonance. (b) Stark-tuned F\"orster resonances observed for various numbers of atoms \textit{i}=1-5 and the initial states $37P_{3/2}(|M|=1/2)$ and $37P_{3/2}(|M|=3/2)$. The two-body resonances are absent for \textit{i}=1, evidencing their two-body nature. The three-body resonances are absent for \textit{i}=1, 2, evidencing their three-body nature. The circles are numerical simulations for \textit{i}=3 discussed in Sec.~IV.B.}
\end{figure}

Figure~1(b) presents the Stark-tuned F\"orster resonances $\rho_i$ observed experimentally in Ref.[20] for various numbers of the interacting atoms \textit{i}=1-5 and initial states $37P_{3/2}(|M|=1/2)$ or $37P_{3/2}(|M|=3/2)$. The signals with \textit{i}=1 do not show neither two-body nor three-body resonances, since there is no interaction at all. The signals with \textit{i}=2 show only the peaks at 1.79 V/cm or 2.0 V/cm, which are the ordinary two-body resonances that occur for all \textit{i}=2-5 and correspond to intersections 3 or 6 in Fig.~1(a). The additional peaks at 1.71~V/cm or 2.14~V/cm are the predicted three-body resonances 1 and 8 of Fig.~1(a) that are absent for \textit{i}=2 and appear only for \textit{i}=3-5. The two-body and three-body peak positions well agree with those in Fig.~1(a). The circles are numerical simulations for \textit{i}=3 discussed in Sec.~IV.B.

Figure~1(b) shows that the two-body and three-body resonances partially overlap. This overlapping increases as \textit{i} grows due to the increase of the total interaction energy and broadening of the two-body resonance. The overlapping can be reduced if a lower Rydberg state is used, and in our paper [20] we demonstrated this for the initial state $36P_{3/2}$. However, the dipole moments of Rydberg atoms scale as $n^2$, and the interaction becomes weaker for the lower states.

Therefore, in this paper we are aimed at finding theoretically the conditions when the three-body resonance is much narrower and better separated from the two-body resonance. This requires to perform a theoretical analysis of three-body resonances for well localized Rydberg atoms in various spatial configurations, as we did for two-body resonances in our earlier theoretical paper [25].

\section{Analytical model for three frozen Rydberg atoms in a triangle spatial configuration}

\subsection{Three frozen Rydberg atoms in a triangle spatial configuration}

In our earlier experiments [4,26], we used only atoms in the initial state $37P_{3/2}(|M|=1/2)$. Therefore, in the related theoretical analysis [25,26] we considered only the two-body resonance 3 of Fig.~1(a) and ignored the possibility of the three-body resonance 1. Our recent experiment [20] has revealed that some atoms undergo a nonresonant transition from the initial state $37P_{3/2}(|M|=1/2)$ to another Stark sublevel $37P_{3/2}(|M^*|=3/2)$ (see Fig.~1), but such a transition is not described by the two-body operator of dipole-dipole interaction. This requires a new theoretical model to be developed. It is a rather complicated problem, since we should take into account all Stark and magnetic sublevels of the interacting Rydberg atoms. Therefore, we will first consider a simplified analytical model for three frozen Rydberg atoms in an equilateral triangle configuration, when the interaction energy for each atom pair is equal.

For three Rydberg atoms in the initial state $37P_{3/2}(|M|=1/2)$, the two F\"orster resonances 1 and 3 of Fig.~1(a) are possible. The three-body resonance 1 corresponds to the resonant transition between collective states $3\times 37P_{3/2}(|M|=1/2) \rightarrow 37S_{1/2}+38S_{1/2}+37P_{3/2}(|M^*|=3/2)$. This transition is, in fact, composed of the two nonresonant two-body relay transitions $3\times 37P_{3/2}(|M|=1/2) \rightarrow 37S_{1/2}+38S_{1/2}+37P_{3/2}(|M|=1/2)\rightarrow 37S_{1/2}+38S_{1/2}+37P_{3/2}(|M^*|=3/2)$ occurring simultaneously. The latter occurs due to the nonresonant exchange interaction $nP_{3/2}(M)+n'S \rightarrow n'S+nP_{3/2}(M^*)$ corresponding to the excitation hopping between \textit{S} and \textit{P} Rydberg atoms [21,25]. Despite the use of a relay, the transfer occurs in a single step, implying a Borromean character of the relay atom, which absorbs the energy of the finite F\"orster defect.

Figure 2 shows a simplified scheme of the Borromean three-body F\"orster resonance $3\times 37P_{3/2}(|M|=1/2) \rightarrow 37S_{1/2}+38S_{1/2}+37P_{3/2}(|M^*|=3/2)$ for three interacting Rydberg atoms. The initially populated collective state 1 is ${\rm 3}\times 37P_{3/2} (|M|=1/2)$. The intermediate collective state 2 is $37S_{1/2} +38S_{1/2} +37P_{3/2} (|M|=1/2)$ with the initial moment projection of the \textit{P} state. The final collective state 3 is $37S_{1/2} +38S_{1/2} +37P_{3/2} (|M^{*}|=3/2)$ with the changed moment projection of the \textit{P} state. The energy defects $\Delta _{1} $ and $\Delta _{2} $ are controlled by the dc electric field. The value of $\Delta _{1} $ can be arbitrary, while $\Delta _{2} $ is nonzero and nearly constant in the vicinity of the F\"orster resonance (being just the Stark splitting between the $|M|=1/2$ and $|M|=3/2$ sublevels). The three-body resonance occurs at $\Delta _{1} =\Delta _{2} $, while two-body resonance occurs at $\Delta _{1} =0$.

\begin{figure}
\includegraphics[scale=0.8]{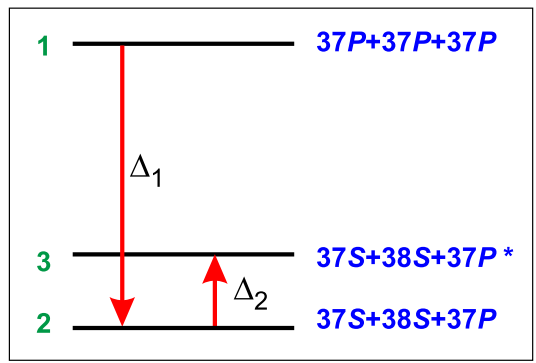}
\caption{\label{Fig2} Simplified scheme of the three-body F\"orster resonance ${\rm 3}\times 37P_{3/2} (|{M}|=1/2)\to 37S_{1/2} +38S_{1/2} +37P_{3/2} (|{M^{*}}|=3/2)$ for three Rydberg atoms. The initially populated collective state 1 is ${\rm 3}\times 37P_{3/2} (|{M}|=1/2)$. The final collective state 3 is $37S_{1/2} +38S_{1/2} +37P_{3/2} (|{M^{*}}|=3/2)$ with the changed moment projection of the \textit{P} state. The intermediate collective state 2 is $37S_{1/2} +38S_{1/2} +37P_{3/2} (|{M}|=1/2)$ with the initial moment projection of the \textit{P} state. The energy defects $\Delta _{1} $ and $\Delta _{2} $ are controlled by the dc electric field. The value of $\Delta _{1} $ can be arbitrary, while $\Delta _{2} $ is nonzero and nearly constant in the vicinity of the F\"orster resonance. The three-body resonance occurs at $\Delta _{1} =\Delta _{2} $, while the two-body one occurs at $\Delta _{1} =0$.}
\end{figure}

For the simplicity, we consider an equilateral-triangle spatial configuration of the three frozen Rydberg atoms spaced by the equal distances \textit{R}. The dipole-dipole matrix element for the two-body transition 1$\rightarrow$2 is given by

\begin{equation} \label{Eq1} 
V_1=\frac{d_{1} d_{2} }{4\pi \varepsilon _{0} } \left[\frac{1}{R^{3} } -\frac{3\, \, Z^{2} }{R^{5} } \right],  
\end{equation} 

\noindent where $d_{1} ,d_{2} $ are the \textit{z} components of the matrix elements of dipole moments of transitions ${\left| 37P_{3/2} \left(M=1/2\right) \right\rangle} \to {\left| 37S_{1/2} \left(M=1/2\right) \right\rangle} $ and ${\left| 37P_{3/2} \left(M=1/2\right) \right\rangle} \to {\left| 38S_{1/2} \left(M=1/2\right) \right\rangle} $, $Z$ is the \textit{z }component of the vector connecting the two atoms $R$ (\textit{z} axis is chosen along the dc electric field), and $\varepsilon_0$ is the dielectric constant. 

The dipole-dipole matrix element $V_2 $ for the two-body transition 2$\rightarrow$3 is given by an equation similar to Eq.~\eqref{Eq1}, but for the \textit{x} and \textit{y} components of the connecting vector and dipole moments of transitions ${\left| 37P_{3/2} \left(M=3/2\right) \right\rangle} \to {\left| 37S_{1/2} \left(M=1/2\right) \right\rangle} $ and ${\left| 37P_{3/2} \left(M=3/2\right) \right\rangle} \to {\left| 38S_{1/2} \left(M=1/2\right) \right\rangle} $.

\subsection{Analytical model}

States 2 and 3 are actually six-fold degenerate with respect to atom permutations. Therefore, there are totally six transitions with the matrix element $V_1$ from state 1 to state 2 and twelve allowed transitions with the matrix element $V_2$ from state 2 to state 3. In addition, the degenerate sublevels of states 2 and 3 experience always-resonant exchange interactions due to the excitation hopping between \textit{S} and \textit{P} Rydberg atoms [21,25]. The Schr\"odinger equation then gives for the probability amplitudes $a_{1} -a_{3} $ of the degenerate sublevels of states 1-3:

\begin{equation} \label{Eq2} 
\begin{array}{l} {i\dot{a}_{1} =6\Omega _{1} a_{2} {\rm e}^{-i\Delta _{1} t} }, \\ \\ {i\dot{a}_{2} =2\Omega _{1} a_{2} +\Omega _{1} a_{1} {\rm e}^{i\Delta _{1} t} +2\Omega _{2} a_{3} {\rm e}^{i\Delta _{2} t} }, \\ \\ {i\dot{a}_{3} =2\Omega _{2} a_{3} +2\Omega _{2} a_{2} {\rm e}^{-i\Delta _{2} t} }. \end{array}   
\end{equation} 

\noindent Here $\Omega_1 =V_1/\hbar $ and $\Omega_2 =V_2 /\hbar $. The terms without exponents in the right-hand sides are responsible for the always-resonant exchange interactions that just shift the energies of states 2 and 3, while the terms with the exponents drive the transitions between collective states.

Equations \eqref{Eq2} can be solved analytically for the arbitrary interaction energy, detunings, and \textit{t} (see Appendix~A). Taking into account the six-fold level degeneracy, the three-atom resonance spectrum is then calculated as

\begin{equation} \label{Eq3} 
\rho _{3} =(6|a_{2} |^{2} +6|a_{3} |^{2} )/3\,. 
\end{equation} 

\noindent This value corresponds to the probability to find one of the three atoms in the final $37S_{1/2} $ state and it is the signal measured in our experiments.

\subsection{Weak dipole-dipole interaction}

Exact analytical solutions for Eqs.~\eqref{Eq2} and \eqref{Eq3} are rather complex and cannot be presented in a clearly understandable way. In this subsection we consider the case of the weak dipole-dipole interaction with $\Omega_1 t, \Omega_2 t\ll 1$, when most of the population remains in the initial state 1, and final states 2 and 3 are weakly populated. In this case, we can set $a_1 \approx 1$, and the approximate analytical solution of Eqs.~\eqref{Eq2} and \eqref{Eq3} yields

\begin{equation} \label{Eq4} 
\begin{array}{c} {\rho _{3} \approx \displaystyle \frac{8\Omega_1 ^{2} }{\Delta _{1}^{2} } \sin ^{2} \left[\frac{\Delta _{1} t}{2} \right]+32\Omega_1 ^{2} \Omega_2 ^{2} \times } \\ \\ {\left\{\displaystyle \frac{1}{\Delta _{1} \Delta _{2} (\Delta _{1} -\Delta _{2} )^{2} } \sin ^{2} \left[\frac{(\Delta _{1} -\Delta _{2} )t}{2} \right]+\right. } \\ \\ {\displaystyle \frac{1}{\Delta _{1} \Delta _{2}^{2} (\Delta _{1} -\Delta _{2} )} \sin ^{2} \left[\frac{\Delta _{2} t}{2} \right]-} \\ \\ {\left. \displaystyle \frac{1}{\Delta _{1}^{2} \Delta _{2} (\Delta _{1} -\Delta _{2} )} \sin ^{2} \left[\frac{\Delta _{1} t}{2} \right]\right\}\,.} \end{array} 
\end{equation} 

The first term in Eq.~\eqref{Eq4} is responsible for the two-body resonance at $\Delta _{1} =0$. The resonance amplitude grows as $\rho _{3} \to 2(\Omega_1 t)^{2} $, while its width is given by the Fourier width of the interaction pulse. Upon spatial averaging, this resonance obtains the cusped line shape, while the Rabi-like population oscillations are washed out, as discussed in Refs.~[26-28].

The third and the fourth terms in Eq.~\eqref{Eq4} do not give any resonance at $\Delta _{1} =\Delta _{2} $ as they compensate for each other with $\Delta _{2} $ being nonzero and almost constant near the F\"orster resonance.  At $\Delta _{1} =0$ the fourth term just reduces the first two-body term, while the second and the third terms compensate for each other as well. 

It is the second term in Eq.~\eqref{Eq4} which is responsible for the Borromean three-body resonance at $\Delta _{1} =\Delta _{2} $. Its amplitude grows as $\rho _{3} \to 8(\Omega_1 \Omega_2 t/\Delta _{2} )^{2} $. The relationship between amplitudes of the three-body and two-body resonances is $(2\Omega_2 /\Delta _{2} )^{2} $. Therefore, the three-body resonance is always weaker than the two-body resonance for the weak dipole-dipole interaction. For example, $\Delta_2/(2\pi)$=9.5~MHz for the $37P_{3/2}(|M|=1/2)$ and $37P_{3/2}(|M|=3/2)$ Stark sublevels of Rb atoms in the electric field of 1.71~V/cm corresponding to the three-body resonance, while the average dipole-dipole interaction energy is on the order of 1~MHz in our experiments.

However, when the three-body resonance is exactly tuned ($\Delta _{1} =\Delta _{2} $), its contribution to the population transfer generally exceeds the contribution from two-body interaction, which is off-resonant in this case. From Eq.~\eqref{Eq4} we find that the three-body contribution relates to the two-body contribution as $2(\Omega_2 t)^{2} $ and can be large for the long interaction times. This is the main condition for the three-body resonance to be of the Borromean type. 

Equation~\eqref{Eq4} helps to understand which parameters are responsible for the two- and three-body F\"orster resonances. It also shows that coherent Rabi-like population oscillations are possible at the exact three-body F\"orster resonance ($\Delta _{1} =\Delta _{2} $) in this simplified model considering equal interaction energies for each atom pair.

\subsection{Strong dipole-dipole interaction}

Equation \eqref{Eq4} is not valid to describe two- and three-body F\"orster resonances at the strong dipole-dipole interaction or long interaction time, since these resonances should saturate and broaden. However, we can consider an approximate solution for the case of three-body F\"orster resonance when $\Delta _{1} $ is scanned in the vicinity of $\Delta _{2} $ ($|\Delta_1-\Delta_2|\ll |\Delta_2|$) and the interaction energies $\Omega_1 ,\Omega_2$ are less than $\Delta _{2} $. In this case the intermediate state 2 of Fig.~2 is almost unpopulated, and the calculated line shape of three-body resonance is given by the formula

\begin{equation} \label{Eq5} 
\rho _{3} \approx \frac{\Omega _{0}^{2} /3}{\left(\Delta -\Delta _{0} \right)^{2} +\Omega _{0}^{2} } \sin ^{2} \left[\frac{t}{2} \sqrt{\left(\Delta -\Delta _{0} \right)^{2} +\Omega _{0}^{2} } \right], 
\end{equation} 

\noindent where $\Delta =\Delta _{1} -\Delta _{2} $ is the detuning from the unperturbed three-body resonance, $\Delta _{0} =-2\Omega _{2} +(4\Omega _{2}^{2} -6\Omega _{1}^{2} )/(\Delta _{2} +2\Omega _{1} )$ is the interaction-induced shift of the three-body resonance, and $\Omega _{0} =4\sqrt{6} \Omega _{1} \Omega _{2} /(\Delta _{2} +2\Omega _{1} )$ is the Rabi-like oscillation frequency. Equation \eqref{Eq5} reveals several important features of the three-body F\"orster resonances. 

First, the resonance experiences the shift $\Delta_0$, which is composed of two parts:  the part with $-2\Omega_2$ is due to the always-resonant exchange interactions, and the other part is the ac dynamic Stark shift induced by the off-resonant dipole-dipole interactions. The three-body resonance  position thus depends on the interaction strength and on the relationship of the dipole-dipole matrix elements at the transitions $1\to 2$ and $2\to 3$ of Fig.~2. In a real Rydberg atom there are the Stark and Zeeman sublevels, which actually open up many interaction channels with different dipole-dipole matrix elements. This can lead to the formation of multiple three-body F\"orster resonances at different resonant electric fields. On the one hand, this complicates the analysis of such resonances due to their possible overlapping. On the other hand, the dynamic shift can separate different interaction channels and provide their maximum coherence for quantum information processing.

Second, Eq.~\eqref{Eq5} shows that coherent Rabi-like population oscillations really take place also at the strong dipole-dipole interaction regime. At the exact resonance ($\Delta =\Delta _{0} $), the Rabi-like oscillation frequency is $\Omega _{0}$, which depends on the interaction strength and on the particular interaction channel. The maximum height of the resonance is 1/3 (one of the three atoms is found to be in the final $37S_{1/2}$ state). The resonance saturates and broadens when the interaction strength increases. The resonance width is determined by a combination of the Fourier width of the interaction pulse and of the three-body interaction strength $\Omega _{0} $, as it takes place for the two-body resonances analyzed in our paper [26].

Third, Eq.~\eqref{Eq5} demonstrates full analogy with a two-photon transition in a three-level system with the far-detuned intermediate state. The intermediate state is not populated and Rabi-like population oscillations occur only between the initial and final states. The three-body oscillation frequency $\Omega _{0} $ is much less than the oscillation frequency of the intermediate two-body resonances ($\Omega_1$ and $\Omega_2$) due to large detuning $\Delta _{2} $. Therefore, the coherence time of the three-body resonance can be much longer than the coherence time of the two-body resonances for the same number of the population oscillations.

Fourth, each Rabi-like oscillation minimum corresponds to a $\pi $ phase shift of the collective wave function of the three interacting Rydberg atoms [29]. As such oscillations are controllable and reversible, they can be used to implement three-qubit quantum gates with Rydberg atoms, for example, the Toffoli or Fredkin gates [22,23]. For this purpose, the most suitable spatial geometries and interaction channels should be found. This requires to perform numerical simulations in order to account for the orientation of atomic dipoles and the Stark or Zeeman sublevels of real Rydberg states, as described in the next section. Numerical simulations can also provide spatial averaging over fluctuating atom positions [25,26], as this is the case in real experiments with optical-trap arrays of neutral atoms.

\section{Full theory and numerical simulations of the dipole-dipole interaction in a three-atom ensemble}

\subsection{Full three-atom operator of the dipole-dipole interaction}

The electric dipole-dipole interaction between two atoms 1 and 2 is described by the operator 

\begin{equation} \label{Eq6} 
\hat{V}_{dd} =\frac{1}{4\pi \varepsilon _{0} R_{12}^{3} } \left[\hat{\mathbf{d}}_{1} \hat{\mathbf{d}}_{2} -3\left(\hat{\mathbf{d}}_{1} \mathbf{n}_{12} \right)\left(\hat{\mathbf{d}}_{2} \mathbf{n}_{12} \right)\right]. 
\end{equation} 

\noindent Here $R_{12} $ is the interatomic distance,  $\mathbf{n}_{12} =\left(\cos \varphi \sin \theta ,\sin \varphi \sin \theta ,\cos \theta \right)$ is a unit vector in the direction connecting two atoms, and $\hat{\mathbf{d}}_{1} $ and $\hat{\mathbf{d}}_{2} $ are dipole-moment operators for these two atoms. By introducing the components of the dipole operators in the spherical basis [30] for each atom \textit{k} as $\hat{d}_{k,\pm } ={\mp \left(\hat{d}_{k,x} \pm i\hat{d}_{k,y} \right) \mathord{\left/{\vphantom{\mp \left(\hat{d}_{k,x} \pm i\hat{d}_{k,y} \right) \sqrt{2} }}\right.\kern-\nulldelimiterspace} \sqrt{2} } $, we expand the operator of dipole-dipole interaction as follows:

\begin{equation*}  
\begin{array}{l} {\hat{V}_{dd} =\displaystyle \frac{1}{4\pi \varepsilon _{0} R_{12}^3 } \times } \\ {\left[A_{1} \left(\theta \right)\left(\hat{d}_{1+} \hat{d}_{2-} +\hat{d}_{1-} \hat{d}_{2+} +2\hat{d}_{1z} \hat{d}_{2z} \right)+\right. } \\ {A_{2} \left(\theta ,\varphi \right)\left(\hat{d}_{1+} \hat{d}_{2z} -\hat{d}_{1-} \hat{d}_{2z} +\hat{d}_{1z} \hat{d}_{2+} -\hat{d}_{1z} \hat{d}_{2-} \right)+} \\ {A_{3} \left(\theta ,\varphi \right)\left(\hat{d}_{1+} \hat{d}_{2z} +\hat{d}_{1-} \hat{d}_{2z} +\hat{d}_{1z} \hat{d}_{2+} +\hat{d}_{1z} \hat{d}_{2-} \right)+} \\ {A_{4} \left(\theta ,\varphi \right)\left(\hat{d}_{1+} \hat{d}_{2+} +\hat{d}_{1-} \hat{d}_{2-} \right)+} \\ {\left. A_{5} \left(\theta ,\varphi \right)\left(\hat{d}_{1+} \hat{d}_{2+} -\hat{d}_{1-} \hat{d}_{2-} \right)\right]\,.} \end{array} 
\end{equation*} 

\noindent Here the angular pre-factors are

\begin{equation*} 
\begin{array}{l} {A_{1} \left(\theta \right)=\displaystyle \frac{1-3\cos ^{2} \left(\theta \right)}{2} }\,, \\ \\{A_{2} \left(\theta ,\varphi \right)=\displaystyle \frac{3\sin \left(2\theta \right)\cos \left(\varphi \right)}{2\sqrt{2} } }\,, \\ \\ {A_{3} \left(\theta ,\varphi \right)=-i\displaystyle \frac{3\sin \left(2\theta \right)\sin \left(\varphi \right)}{2\sqrt{2} } }\,, \\ \\{A_{4} \left(\theta ,\varphi \right)=-\displaystyle \frac{3\sin ^{2} \left(\theta \right)\cos \left(2\varphi \right)}{2} }\,, \\ \\{A_{5} \left(\theta ,\varphi \right)=i\displaystyle \frac{3\sin ^{2} \left(\theta \right)\sin \left(2\varphi \right)}{2} }\,. \end{array}   
\end{equation*} 

The operator $\hat{V}_{dd} $ couples the states where the total magnetic quantum number $M=M_{1} +M_{2} $ changes by $\Delta M=0, \pm 1, \pm 2$. The matrix element of the $\hat{V}_{dd} $ operator for a transition between the collective two-atom states ${\left| \gamma _{a} ,\gamma _{b}  \right\rangle} \to {\left| \gamma _{s} ,\gamma _{t}  \right\rangle} $,  where for each atomic state ${\left| \gamma  \right\rangle} ={\left| nLJM \right\rangle} $ \textit{n} is the principal quantum number, \textit{L} is the orbital moment,  \textit{J} is the total moment and \textit{M} is the projection of the total moment, is expressed as [31,32]

\begin{equation*} 
\begin{array}{l} {{\left\langle n_{s} M_{s} L_{s} J_{s} ;n_{t} M_{t} L_{t} J_{t}  \right|} \hat{V}_{dd} {\left| n_{a} M_{a} L_{a} J_{a} ;n_{b} M_{b} L_{b} J_{b}  \right\rangle} =} \\ \\{=\displaystyle \frac{e^{2} }{4\pi \varepsilon _{0} R_{12}^{3} } \left\{A_{1} \left(\theta \right)\left[C_{J_{a} M_{a} 11}^{J_{s} M_{s} } C_{J_{b} M_{b} 1-1}^{J_{t} M_{t} } +\right. \right. } \\ \\{\left. C_{J_{a} M_{a} 1-1}^{J_{s} M_{s} } C_{J_{b} M_{b} 11}^{J_{t} M_{t} } +2C_{J_{a} M_{a} 10}^{J_{s} M_{s} } C_{J_{b} M_{b} 10}^{J_{t} M_{t} } \right]+} \\ \\{A_{2} \left(\theta ,\varphi \right)\left[\left(C_{J_{a} M_{a} 11}^{J_{s} M_{s} } -C_{J_{a} M_{a} 1-1}^{J_{s} M_{s} } \right)C_{J_{b} M_{b} 10}^{J_{t} M_{t} } +\right. } \\ \\{\left. C_{J_{a} M_{a} 10}^{J_{s} M_{s} } \left(C_{J_{b} M_{b} 11}^{J_{t} M_{t} } -C_{J_{b} M_{b} 1-1}^{J_{t} M_{t} } \right)\right]+} \\ \\{A_{3} \left(\theta ,\varphi \right)\left[\left(C_{J_{a} M_{a} 11}^{J_{s} M_{s} } +C_{J_{a} M_{a} 1-1}^{J_{s} M_{s} } \right)C_{J_{b} M_{b} 10}^{J_{t} M_{t} } +\right. } \\ \\{\left. C_{J_{a} M_{a} 10}^{J_{s} M_{s} } \left(C_{J_{b} M_{b} 11}^{J_{t} M_{t} } +C_{J_{b} M_{b} 1-1}^{J_{t} M_{t} } \right)\right]+} \\ \\{A_{4} \left(\theta ,\varphi \right)\left[C_{J_{a} M_{a} 11}^{J_{s} M_{s} } C_{J_{b} M_{b} 11}^{J_{t} M_{t} } +C_{J_{a} M_{a} 1-1}^{J_{s} M_{s} } C_{J_{b} M_{b} 1-1}^{J_{t} M_{t} } \right]+} \\ \\{\left. A_{5} \left(\theta ,\varphi \right)\left[C_{J_{a} M_{a} 11}^{J_{s} M_{s} } C_{J_{b} M_{b} 11}^{J_{t} M_{t} } -C_{J_{a} M_{a} 1-1}^{J_{s} M_{s} } C_{J_{b} M_{b} 1-1}^{J_{t} M_{t} } \right]\right\}\times } \\ \\{\sqrt{\max \left(L_{a} ,L_{s} \right)} \sqrt{\max \left(L_{b} ,L_{t} \right)} \sqrt{\left(2J_{a} +1\right)\left(2J_{b} +1\right)} \times } \\ \\{\times \left\{\begin{array}{ccc} {L_{a} } & {1/2} & {J_{a} } \\ \\ {J_{s} } & {1} & {L_{s} } \end{array}\right\}\left\{\begin{array}{ccc} {L_{b} } & {1/2} & {J_{b} } \\ \\ {J_{t} } & {1} & {L_{b} } \end{array}\right\}\left(-1\right)^{L_{s} +\frac{L_{a} +L_{s} +1}{2} } \times } \\ \\ {\left(-1\right)^{L_{t} +\frac{L_{b} +L_{t} +1}{2} } \left(-1\right)^{J_{a} +J_{b} } R_{n_{a} L_{a} }^{n_{s} L_{s} } R_{n_{b} L_{b} }^{n_{t} L_{t} } }\,. \end{array}  
\end{equation*}

\noindent Here $R_{n_{a} L_{a} }^{n_{s} L_{s} } $and $R_{n_{b} L_{b} }^{n_{t} L_{t} } $ are radial matrix elements for ${\left| n_{a} L_{a}  \right\rangle} \to {\left| n_{s} L_{s}  \right\rangle} $ and ${\left| n_{b} L_{b}  \right\rangle} \to {\left| n_{t} L_{t}  \right\rangle} $ transitions, respectively. The radial matrix elements in Rydberg atoms are calculated using the quasiclassical approximation [33].

In our numerical simulations we consider collective states of the three atoms ${\left| \gamma _{1} \gamma _{2} \gamma _{3}  \right\rangle} ={\left| n_{1} L_{1} J_{1} M_{1} ;n_{2} L_{2} J_{2} M_{2} ;n_{3} L_{3} J_{3} M_{3}  \right\rangle} $. For example, if the atoms are initially excited to the $37P_{3/2}$ state, we take into account the eight atomic states ${\left| 37P_{3/2} ,M=\pm {1 \mathord{\left/{\vphantom{1 2}}\right.\kern-\nulldelimiterspace} 2}  \right\rangle} $, ${\left| 37P_{3/2} ,M=\pm {3 \mathord{\left/{\vphantom{3 2}}\right.\kern-\nulldelimiterspace} 2}  \right\rangle} $, ${\left| 37S_{1/2} ,M=\pm {1 \mathord{\left/{\vphantom{1 2}}\right.\kern-\nulldelimiterspace} 2}  \right\rangle} $, and ${\left| 38S_{1/2} ,M=\pm {1 \mathord{\left/{\vphantom{1 2}}\right.\kern-\nulldelimiterspace} 2}  \right\rangle} $. The initial state of the three-atom system is taken as the defined or random superposition of the eight degenerate collective states where all three atoms are in the ${\left| 37P_{3/2} ,M=\pm {1 \mathord{\left/{\vphantom{1 2}}\right.\kern-\nulldelimiterspace} 2}  \right\rangle} $ states (or in the ${\left| 37P_{3/2} ,M=\pm {3 \mathord{\left/{\vphantom{3 2}}\right.\kern-\nulldelimiterspace} 2}  \right\rangle} $ states, if the exciting laser is tuned to the excitation of the $37P_{3/2}$ atoms with $\left|M\right|={3 \mathord{\left/{\vphantom{3 2}}\right.\kern-\nulldelimiterspace} 2} $), with the defined or equal statistical weights. For the random superposition, we assume that after laser excitation the sign of $M=M_{1} +M_{2} $ for the initial state is undetermined. The F\"orster energy defect is the difference between the energies of the final collective state  ${\left| \gamma _{1} \gamma _{2} \gamma _{3}  \right\rangle} $ and of the initial state. 

To reduce the complexity of the calculations, we do not take into account far-detuned collective states with F\"orster energy defect exceeding 2~GHz. There are totally 160 collective Rydberg states involved in the calculations with 160 equations describing all possible allowed interactions between collective states.

\subsection{Numerical simulations for three disordered Rydberg atoms in a single interaction volume}

In Ref.[20] we simulated the experimental data of Fig.~1(b), using the method described in Ref.~[25]. It is based upon solving the Schr\"odinger equation with subsequent Monte Carlo averaging over the random positions of the three atoms in a single interaction volume. For this purpose, we considered the three atoms, randomly located in a single cubic interaction volume with the edge length \textit{d}, as shown in Fig.~3. The quantization axis \textit{z} was chosen along the controlling external electric field. For each random spatial configuration we calculated the interatomic distances $R_{12} $, $R_{23} $, $R_{13} $ and the angles $\theta _{12} $, $\theta _{23} $, $\theta _{13} $, $\varphi _{12} $, $\varphi _{23} $, $\varphi _{13} $  between the quantization axis and the vectors connecting the atoms. 

\begin{figure}
\includegraphics[scale=0.35]{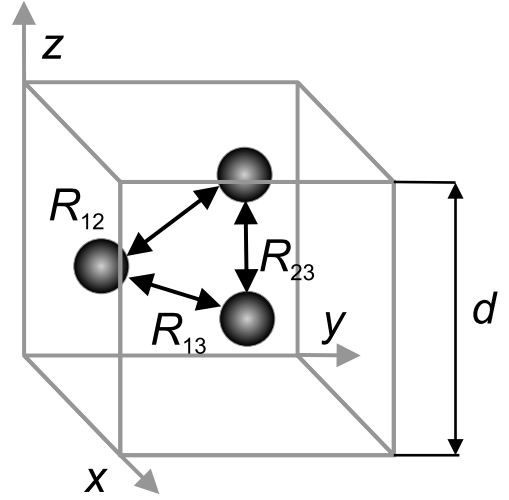}
\caption{\label{Fig3} Geometry of the interaction of the three Rydberg atoms. The atoms are positioned either randomly in a cubic volume to perform numerical simulations with Monte Carlo averaging over the atom positions, or are spatially fixed to provide Rabi-like population oscillations.}
\end{figure}

We then calculated the matrix of the Hamiltonian for collective states ${\left| \gamma _{1} \gamma _{2} \gamma _{3}  \right\rangle} $ of the three-atom system, taking into account the Stark shifts of the atomic energy levels in the controlling dc electric field as the variation of the diagonal terms of the Hamiltonian and the dipole-dipole interaction of atoms. Then we solved numerically the Schr\"odinger equation for the probability amplitudes of all collective states.  The probability $\rho _{3} $ to find one of the atoms in the final $37S$ state for initially excited $37P_{3/2}$ atoms was calculated versus the controlling dc electric field. For each field value we averaged the calculated probabilities over 1000 random spatial configurations. 

The numerical results are presented as the circles in Fig.~1(b). The theoretical spectra were calculated for the cubic interaction volume of 15$\times$15$\times$15 $\mu$m$^3$ and interaction time of 3 $\mu$s which correspond to our experimental parameters. The overall agreement of the full theory with the experiment is satisfactory. The calculated line shapes of the two-body resonances are close to the experimental ones. These are cusp-shaped resonances that are formed upon spatial averaging in a single interaction volume [26-28]. The three-body resonances are also well reproduced by theory, in both their heights and widths. 

\begin{figure*}
\includegraphics[scale=0.9]{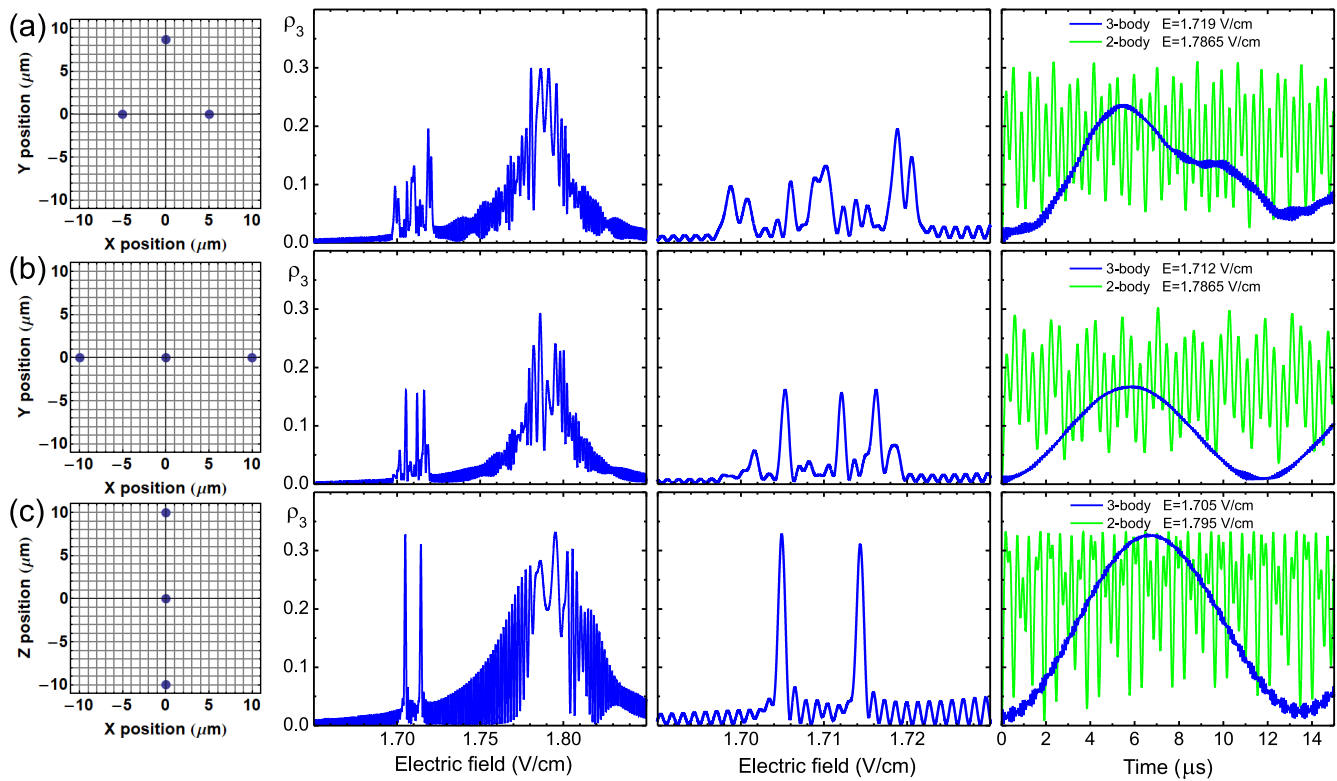}
\caption{\label{Fig4} Numerically calculated three-atom Stark-tuned F\"orster resonances ${\rm 3}\times 37P_{3/2} (M=+1/2)\to 37S_{1/2} +38S_{1/2} +37P_{3/2} (M=\pm 3/2)$ in Rb Rydberg atoms for the three spatial configurations: (a) equilateral triangle in the \textit{x-y} plane; (b) one-dimensional chain along \textit{x} axis; (c) one-dimensional chain along \textit{z} axis. The \textit{z} axis is directed along the controlling electric field. The three atoms are supposed to be completely immobile. The left panels show the corresponding spatial configurations with the micron-sized grids. The second from the left panels are the spectra of three-atom F\"orster resonances calculated for the interaction time of 7 $\mu $s. The broad saturated two-body resonance is centered near 1.79 V/cm, while there are several narrow three-body resonances near 1.71 V/cm. The third from the left panels show the same three-body resonances zoomed in to demonstrate the Rabi-like population oscillations. The right panels show the Rabi-like population oscillations for the centers of the two-body resonance [the green (light gray) curves] and of the most intense three-body resonance [the blue (dark gray) curves]. The narrow three-body resonances require precise setting of the resonant electric field that depends on the spatial configuration and interaction channel.}
\end{figure*}

However, from Fig.~1(b) we see that for the disordered atoms, the three-body resonances are much weaker than the two-body ones, and their width is too large to provide coherent population oscillations predicted by Eq.~\eqref{Eq5} for the frozen atoms in an equilateral triangle configuration. We conclude that one needs to localize the three-atoms in space in order to fix the interaction energies for each pair of atoms and provide the coherence.

\subsection{Numerical simulations for three spatially fixed Rydberg atoms}

Figure 4 presents the numerically calculated three-atom Stark-tuned F\"orster resonances ${\rm 3}\times 37P_{3/2} (M=+1/2)\to $ $37S_{1/2} +38S_{1/2} +37P_{3/2} (M=\pm 3/2)$ in Rb Rydberg atoms for the three spatial configurations: (a) equilateral triangle in the \textit{x-y} plane; (b) one-dimensional chain along \textit{x} axis; (c) one-dimensional chain along \textit{z} axis. The \textit{z} axis is directed along the controlling electric field. The three atoms are supposed to be completely immobile and initially excited to the defined state $37P_{3/2} (M=+1/2)$. The left panels show the corresponding spatial configurations with the micron-sized grids. The total interaction energy for three atoms is nearly the same in all configurations. The second from the left panels are the spectra of three-atom F\"orster resonances calculated for the interaction time of 7~$\mu $s. This time is near the maximum of three-body population transfer at the interatomic spacing used (\textit{R}=10~$\mu $m). The broad saturated two-body resonance is centered near 1.79~V/cm, while there are several narrow three-body resonances near 1.71~V/cm. These resonances correspond to the different three-body interaction channels with different dipole-dipole matrix elements $\Omega_1$ and $\Omega_2$ and different dynamic shifts $\Delta _{0} =-2\Omega _{2} +(4\Omega _{2}^{2} -6\Omega _{1}^{2} )/(\Delta _{2} +2\Omega _{1} )$, as discussed in Section III. The third panels from the left show the same three-body resonances zoomed in to demonstrate the Rabi-like population oscillations at the wings and related coherence. The right panels show the Rabi-like population oscillations for the centers of the two-body resonance [green (light gray) curves] and of the most intense three-body resonance [blue (dark gray) curves]. The narrow three-body resonance requires precise setting of the resonant electric field that depends on the spatial configuration and interaction channel.

It is quite surprising that the symmetrical triangle configuration of Fig.~4(a) with (presumably) equal interaction energy for each atom pair delivers the most complicated structure of the three-body resonance. There are about six partially overlapped resonances, each representing a different interaction channel with its own energy and dynamic shift. All of these resonances occur as the dipole-dipole interaction drives all allowed transitions with $\Delta M=0, \pm 1, \pm 2$. The amplitudes of the three-body resonances do not reach their maximum possible value of 1/3 due to population leaking between these channels. Therefore, the three-body Rabi-like population oscillations do not demonstrate full coherence in this spatial configuration. The two-body population oscillations are also partially dephased and demonstrate irregular character due to several interaction channels involved. The Rabi frequencies are significantly different for the two- and three-body resonances, because the three-body resonance is a weaker second-order relay process occurring via an intermediate state, as discussed in Section~III. 

The linear chain along \textit{x} axis delivers a simpler structure of the three-body resonance in Fig.~4(b). There are 3-4 peaks that are better resolved than in Fig.~4(a). The linear spatial configuration is distinguished by the fact that the central atom interacts with the two side atoms, while each side atom interacts mainly with the central atom. This weakens some of the possible interaction channels, because at the beginning of the three-body interaction the central atom turns out to be most likely in the final $37P_{3/2} (M=\pm 3/2)$ states, while the side atoms are mainly in the final $37S_{1/2}$ or $38S_{1/2}$ states. Nevertheless, at the end of the three-body transition these states are mixed by the always-resonant exchange interaction, and we see again that the amplitudes of the three-body resonances do not reach their maximum possible value of 1/3, and the dephasing of the two-body population oscillations is also present. 

We have finally found that only the linear chain along \textit{z} axis of Fig.~4(c) provides full coherence of the three-body resonances. Compared to Fig.~4(b), this linear chain closes more interaction channels due to the specific selection rule for the change of the moment projection $\Delta M=0$ in this particular spatial configuration, with the dipole-dipole interactions described by Eq.~(1). As a result, there appear only two well-resolved three-body resonances, although they do not correspond to the single final collective states but rather to the coherent superposition of several ones. Figure~4(c) demonstrates high coherence of the population oscillations at the three-body resonance, as their amplitude now reaches 1/3. The coherence of the two-body resonance also increases. The left wing of the two-body resonance contributes only about 5\% to the amplitude of the three-body resonances. This contribution can be made even smaller for longer interaction times and larger atom separations. This means that the three-body resonance is of the Borromean type and represents a high-fidelity three-body operator that indeed can be used in quantum simulations and quantum gates. 

\begin{figure*}
\includegraphics[scale=1]{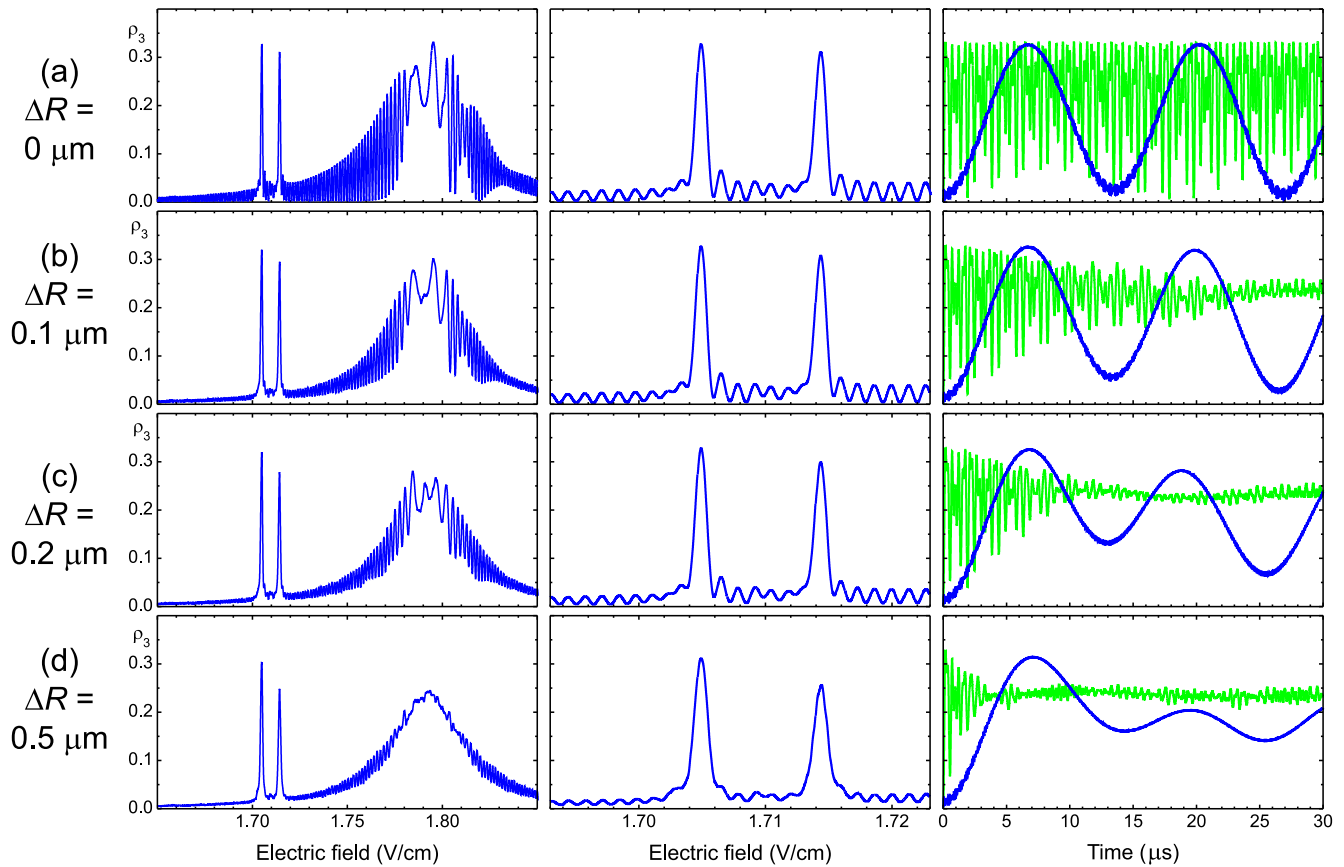}
\caption{\label{Fig5} Numerically calculated three-atom Stark-tuned F\"orster resonances ${\rm 3}\times 37P_{3/2} (M=+1/2)\to 37S_{1/2} +38S_{1/2} +37P_{3/2} (M=\pm 3/2)$ in Rb Rydberg atoms for the spatial configuration of one-dimensional chain along \textit{z} axis with the atom spacing \textit{R}=10 $\mu $m and uncertainty in the position of each atom: (a)\textit{ }$\Delta R$=0 $\mu $m; (b) $\Delta R$=0.1 $\mu $m; (c) $\Delta R$=0.2 $\mu $m; (d)~$\Delta R$=0.5~$\mu $m. The left panels show the spectra of three-atom F\"orster resonances calculated for the interaction time of 7 $\mu $s. The middle panels show the three-body resonance zoomed in to demonstrate the Rabi-like population oscillations. The right panels show the Rabi-like population oscillations for the centers of the two-body resonance at 1.795 V/cm [the green (light gray) curves] and of the three-body resonance at 1.705 V/cm [the blue (dark gray) curves].}
\end{figure*}

For the interaction time of 7 $\mu$s used in Fig.~4(c), the width of the three-body resonances is extremely small ($\sim$1 mV/cm, or $\sim$0.13 MHz in the frequency scale) and is limited by the Fourier-transform width of the interaction pulse, in agreement with Eq.~\eqref{Eq5}. At the same time, the two-body resonance is saturated and broadened. Its width ($\sim$40 mV/cm, or $\sim$5 MHz in the frequency scale) corresponds mainly to the two-body interaction energy, as discussed in our paper [26].

We note that numerical simulations in Fig.~4(c) have been performed for immobile atoms, while in a realistic experiment (e.g., with microscopic optical-trap arrays [34,35]) the atom positions fluctuate, typically within 0.1-1 $\mu $m. Therefore, the next issue to analyze is the effect of the atom-position fluctuations on the coherence of three-body Rabi-like population oscillations and their possible dephasing.

\subsection{Effect of the spatial fluctuations of atoms}

Figure 5 presents the numerically calculated three-atom Stark-tuned F\"orster resonances ${\rm 3}\times 37P_{3/2} (M=+1/2)\to 37S_{1/2} +38S_{1/2} +37P_{3/2} (M=\pm 3/2)$ in Rb Rydberg atoms for the spatial configuration of the linear chain along \textit{z} axis with the atom spacing \textit{R}=10 $\mu $m and uncertainty in the position of each atom: (a) $\Delta R$=0~$\mu $m; (b) $\Delta R$=0.1 $\mu $m; (c) $\Delta R$=0.2 $\mu $m; (d) $\Delta R$=0.5 $\mu $m averaged over 100 random atom positions. The left panels show the spectra of three-atom F\"orster resonances calculated for the interaction time of 7 $\mu $s. The middle panels show the three-body resonances zoomed in to demonstrate the Rabi-like population oscillations at the wings. The right panels show the Rabi-like population oscillations for the centers of the two-body resonance at 1.795~V/cm [the green (light gray) curves] and of the  three-body resonance at 1.705 V/cm [the blue (dark gray) curves] over an extended time scale of 30  $\mu$s.

We see that even the small uncertainty $\Delta R$=0.1 $\mu $m in Fig.~5(b) noticeably affects the coherence of the two-body resonance for times exceeding 5~$\mu $s and its fast Rabi-like population oscillations are partly dephased and washed out in the spectrum and in the time dependence. At the same time, the dephasing of the three-body resonance is weaker due to its slower time dynamics, according to Eq.~\eqref{Eq5}. At 7 $\mu$s, the height of the tree-body resonance is nearly the same as in Fig.~5(a) where the position fluctuations are absent. This means that, in spite of the two-body decoherence, we still have coherent three-body interactions at 7 $\mu$s and therefore can perform three-qubit quantum gates and simulations on this time scale.

However, for the longer interaction times in Fig.~5(b), the three-body oscillations experience visible dephasing for the first oscillation minimum at 13~$\mu $s. This dephasing seems to have some coherent nature due to reversible population transfer to the other collective states, since the second oscillation minimum at 26~$\mu $s turns out to be deeper than the first minimum. This observation is also true for Fig.~5(a) where the three atoms are frozen. Therefore, the three-body oscillations have much longer coherence time than the two-body ones.

The longer coherence time of the three-body resonances does not necessarily imply that they are better suited to quantum gates than the two-body resonances. We should actually consider the relationship between the number of the observed Rabi-like population oscillations and their period. This relationship estimates the number of quantum gates, which can be performed during the coherence time, and in Figs.~5(a) and 5(b) it is better for the two-body resonances. Nevertheless, the three-body resonances are of interest as they provide a platform for the three-qubit quantum gates and simulations, which are inaccessible with the two-body resonances. In addition, three-qubit gates can be decomposed as a sequence of six two-qubit gates, so that overall performance of the two-qubit and three-qubit quantum computations with F\"orster resonances can be similar.

For the larger atom position uncertainties of Figs. 5(c) and 5(d), the dephasing becomes significant. The heights of the two- and three-body resonances decrease and their coherence becomes unacceptably low. The three-body resonance also broadens in Fig.~5(d) due to fluctuations of the dynamic shift. 

Therefore, when implementing the three-qubit quantum gates with the three-body F\"orster resonances, one needs to cool the atoms in the three optical dipole traps down to their vibrational ground states and localize them with the uncertainty of less than 1\% with respect to the interatomic separation. This can be challenging since additional laser cooling stage is required (for example, sideband Raman cooling to the vibrational ground state). On the other hand, additional cooling is also desirable to improve the fidelity of the two-qubit quantum gates, which still does not exceed 80\% in the experiments with Rydberg atoms, presumably due to the residual Doppler effect and motional decoherence of the atoms in optical dipole traps [5,36].

\subsection{Effect of the finite Rydberg lifetimes}

All numerical calculations presented above were done with the Schr\"odinger equation, which does not take into account finite radiative lifetimes of the Rydberg states. Although the lifetimes are typically long and grow as $n^3$ [1], they can contribute to the dephasing of Rabi-like population oscillations at F\"orster resonances. In addition, Rydberg lifetimes are also reduced by the background blackbody radiation. For example, at the ambient temperature of 300~K, the calculated lifetimes of our Rydberg states $37P$, $37S$ and $38S$ are 41, 29 and 31 $\mu$s, correspondingly [37]. These values are comparable with the period of three-body population oscillations in Fig.~5.

However, the radiative decay of Rydberg states occurs mainly to the low-excited states [1,36], while the three-atom signals, which we measure in experiments,  are the fractions of the atoms that have undergone a transition to the final $37S$ state. Since these atoms decay nearly at the same rate as the atoms in the initial $37P$ state, the signal is automatically normalized on the total number of Rydberg atoms and it is thus insensitive to Rydberg lifetimes. Therefore, the population oscillations of Fig.~5 should be observable experimentally in spite of the finite Rydberg lifetimes, since we post-select the signals with exactly three detected atoms.

In order to take the lifetimes into account in the numerical simulations, we need to use the density-matrix equations, as we did in Ref.~[26] for the analysis of the two-body resonance line shape. In the case of three atoms, however, this is practically impossible for the full Zeeman and Stark structures of the used Rydberg states due to the about 10-fold increase in the number of equations to be solved numerically (several thousand equations).

Finally, although the observed three-body population oscillations should not be affected by Rydberg lifetimes, the finite lifetimes would certainly affect the fidelity of three-qubit quantum gates, as some of the atoms can simply decay to low-excited states and the whole gate will be destroyed. Therefore, when considering the three-qubit gates based on three-body F\"orster resonances, we should choose the interaction times that are much shorter than the radiative lifetimes [29]. For example, if the interaction time is $\sim$2 $\mu$s, the lifetimes should be $\sim$200~$\mu$s, which correspond to the Rydberg states with $n\sim 80$ [37]. For such high Rydberg states the simple three-body F\"orster resonances we have considered in this paper are not available [38]. Nevertheless, the required three-body resonances for the atoms with $n\sim 80$ can be engineered either by using the radiofrequency electric field that creates Floquet sidebands of Rydberg states [38], or by using more complicated F\"orster resonances with the three atoms initially excited to different Rydberg states. 

For example, in our forthcoming paper [29] we consider the Stark-tuned three-body F\"orster resonance for the initial collective state $\vert 80P_{3/2}(M=+3/2); 81P_{3/2}(M=+3/2);81P_{3/2}(M=-3/2)\rangle$ and show that a three-qubit Toffoli gate can be implemented with a fidelity exceeding 96\%. We have also found that addition of a magnetic field ($\sim$ 1 G) reduces the complexity of the three-body F\"orster resonances by lifting the Zeeman degeneracy and and isolating the three-body interaction channels.

\section{Conclusions}

In this paper we theoretically investigated the coherence of the Borromean three-body F\"orster resonances. In particular, we studied if the Rabi-like population oscillations are possible for well localized Rydberg atoms in various spatial configurations, when the interaction energy is well fixed. 

We have first built a simple analytical model for an equilateral-triangle spatial configuration of the three interacting Rydberg atoms and found the approximate formulas for the weak and strong dipole-dipole interaction. They both show that coherent Rabi-like population oscillations are possible at the exact three-body F\"orster resonance. We have found, however, that the three-body resonance experiences a dynamic shift and its position in the electric-field scale depends on the interaction strength and on the relationship of the dipole-dipole matrix elements involved in the particular three-body interaction channel.

We have further built a numerical model in order to account for the orientation of atomic dipoles and the Stark or Zeeman sublevels of real Rydberg states. This model shows that, in general, there appear multiple three-body F\"orster resonances, which correspond to the different three-body interaction channels with different dipole-dipole matrix elements and dynamic shifts. 

It is quite surprising that the spatial configuration of an equilateral triangle in the \textit{x-y} plane (\textit{z} axis is directed along the controlling electric field) with (presumably) equal interaction energy for each atom pair delivers the most complicated structure of the three-body resonance. There are about six partially overlapped resonances, each representing a different interaction channel with its own energy and dynamic shift. The amplitudes of the three-body resonances do not reach their maximum possible value of 1/3 due to the population leaking between these channels. Therefore, the three-body Rabi-like population oscillations do not provide full coherence in this spatial configuration. 

The linear chain along \textit{x} axis delivers a simpler structure of the three-body resonance, because at the beginning of the three-body interaction the central atom turns out to be most likely in the final \textit{P} state, while the side atoms are mainly in the final \textit{S} states. Nevertheless, at the end of the three-body transition these states are mixed by the always-resonant exchange interaction, and the amplitudes of the three-body resonances do not reach their maximum possible value of 1/3.

We have finally revealed that the linear chain along \textit{z} axis provides full coherence of the three-body resonances. This configuration closes some interaction channels due to the specific selection rule for the change of the moment projection in this particular spatial configuration. As a result, there are only two well-resolved three-body resonances that demonstrate high coherence of the population oscillations at the three-body resonance, as their amplitude now reaches 1/3. The two-body resonance has a small contribution (below 5\%) to these resonances. This means that the three-body resonance is of the Borromean type and represents a high-fidelity three-body operator that indeed can be used in quantum simulations and quantum gates. 

Numerical simulations also provided spatial averaging over fluctuating atom positions, as this is the case in real experiments with optical-trap arrays of neutral atoms. We have found that, on the time scale of 10 $\mu$s, even small spatial uncertainty noticeably affects the coherence of the two-body resonance, but the dephasing of the three-body resonance is much weaker due to its much slower time dynamics. This means that, in spite of the two-body decoherence, we still have coherent three-body interactions and therefore can perform the three-qubit quantum gates and simulations. However, for longer interaction times or larger position uncertainties, the three-body oscillations experience noticeable dephasing even at the first oscillation minimum. Therefore, when implementing the three-qubit quantum gates with the three-body F\"orster resonances, one needs to cool the atoms in the three optical dipole traps down to their vibrational ground states and localize them with an uncertainty of less than 1\% with respect to the interatomic separation.

Finally, although the observed three-body population oscillations should not be affected by Rydberg lifetimes, since the measured signals are normalized on the number of atoms, the finite lifetimes would certainly affect the fidelity of three-qubit quantum gates, as some of the atoms can simply decay to low-excited states and the whole gate will be destroyed. Therefore, when considering the three-qubit gates based on three-body F\"orster resonances, we should choose the interaction times that are much shorter than the radiative lifetimes [29].

In this respect we note that each Rabi-like oscillation minimum corresponds to a $\pi $ phase shift of collective wave function of the three interacting Rydberg atoms [29]. If such oscillations are controllable and reversible, they can be used to implement three-qubit quantum gates with Rydberg atoms, for example, the Toffoli or Fredkin gates, which provide further speed up of quantum computation and implementation of quantum error-correction algorithms [22,23]. Coherent electrically-controlled three-body interactions can also be used in quantum simulators, e.g., to implement the Bose-Hubbard model, where the transition from a superfluid state to the Mott insulator state is modified by such interaction [39], in realization of the fractional quantum Hall effect based on three-body interactions [40], or in implementation of the topological insulators [41]. 
\\

\begin{acknowledgments}
The authors are grateful to Elena Kuznetsova for fruitful discussions. This work was supported by the Russian Science Foundation Grant No 16-12-0083 in the part of the numerical model and Grant No 18-12-00313 in the part of the analytical calculations, the RFBR Grant No 17-02-00987 in the part of simulation of the off-resonant Rydberg interactions and Grant No 16-02-00383 in part of simulation of coherent three-body F\"orster resonances, the Novosibirsk State University, the public Grant CYRAQS from Labex PALM (ANR-10-LABX-0039) and the EU H2020 FET Proactive project RySQ (Grant No.~640378).
\end{acknowledgments}

\section*{APPENDIX A. Analytical solution for a Borromean three-body F\"orster resonance}

For the simplified scheme of the Borromean three-body F\"orster resonance ${\rm 3}\times nP_{3/2} (|M|=1/2)\to nS_{1/2} +(n+1)S_{1/2} +nP_{3/2} (|M^*|=3/2)$ shown in Fig.~2, the Schr\"odinger equation gives for the probability amplitudes $a_{1} -a_{3} $ of the degenerate sublevels of states 1-3:

\begin{equation} \label{Eq7}
\begin{array}{l} {i\dot{a}_{1} =6\Omega _{1} a_{2} {\rm e}^{-i\Delta _{1} t} }, \\ \\ {i\dot{a}_{2} =2\Omega _{1} a_{2} +\Omega _{1} a_{1} {\rm e}^{i\Delta _{1} t} +2\Omega _{2} a_{3} {\rm e}^{i\Delta _{2} t} }, \\ \\ {i\dot{a}_{3} =2\Omega _{2} a_{3} +2\Omega _{2} a_{2} {\rm e}^{-i\Delta _{2} t} }. \end{array}  
\end{equation}

\noindent Here $\Omega _{1} =V_{1} /\hbar $ and $\Omega _{2} =V_{2} /\hbar $ are the matrix elements (in the frequency units) of dipole-dipole interactions at the transitions $1\to 2$ and $2\to 3$. The terms without exponents in the right-hand sides are responsible for the always-resonant exchange interactions that just shift the energies of states 2 and 3, while the terms with the exponents drive the transitions between collective states. 

Equations~\eqref{Eq7} can be solved analytically for the arbitrary interaction energy, detunings, and \textit{t}. We first do the replacements

\begin{equation} \label{Eq8}
\begin{array}{l} {a_{2} =\alpha _{2} e^{-2i\Omega _{1} t} }, \\ \\ {a_{3} =\alpha _{3} e^{-2i\Omega _{2} t} }, \end{array}   
\end{equation}

\noindent and obtain a modified Eq.~\eqref{Eq7} as

\begin{equation} \label{Eq9}
\begin{array}{l} {i\dot{a}_{1} =6\Omega _{1} \alpha _{2} {\rm e}^{-i(\Delta _{1} +2\Omega _{1} )t} },\\ \\ {i\dot{\alpha }_{2} =\Omega _{1} a_{1} {\rm e}^{i(\Delta _{1} +2\Omega _{1} )t} +2\Omega _{2} \alpha _{3} {\rm e}^{i(\Delta _{2} +2\Omega _{1} -2\Omega _{2} )t} },\\ \\ {i\dot{\alpha }_{3} =2\Omega _{2} \alpha _{2} {\rm e}^{-i(\Delta _{2} +2\Omega _{1} -2\Omega _{2} )t}. } \end{array} 
\end{equation}

After several substitutions we come to a single differential equation

\begin{equation} \label{Eq10}
\begin{array}{l} {\dddot{\alpha }_{3} +i(2\Delta _{2} -\Delta _{1} +2\Omega _{1} -4\Omega _{2} )\ddot{a}_{3} +} \\ \\{[(\Delta _{2} +2\Omega _{1} -2\Omega _{2} )(\Delta _{1} -\Delta _{2} +2\Omega _{2} )+} \\ \\{6\Omega _{1}^{2} +4\Omega _{2}^{2} ]\dot{\alpha }_{3} -4i(\Delta _{1} -\Delta _{2} +2\Omega _{2} )\Omega _{2}^{2} =0}. \end{array}
\end{equation}

Then we seek the solution as $\alpha _{3} \sim {\rm e}^{i\mu t} $ and obtain the cubic equation

\begin{equation} \label{Eq11}
\begin{array}{l} {\mu ^{3} +(2\Delta _{2} -\Delta _{1} +2\Omega _{1} -4\Omega _{2} )\mu ^{2} -} \\ \\{[(\Delta _{2} +2\Omega _{1} -2\Omega _{2} )(\Delta _{1} -\Delta _{2} +2\Omega _{2} )+} \\ \\{6\Omega _{1}^{2} +4\Omega _{2}^{2} ]\mu +4(\Delta _{1} -\Delta _{2} +2\Omega _{2} )\Omega _{2}^{2} =0}. \end{array}
\end{equation}

\noindent This equation has three roots, which are found analytically with the following sequence of equations taken from the mathematical handbooks:

\begin{equation*}
\begin{array}{l}
{S=-[(\Delta _{2} +2\Omega _{1} -2\Omega _{2} )(\Delta _{1} -\Delta _{2} +2\Omega _{2} )+6\Omega _{1}^{2} +4\Omega _{2}^{2} ]\,,} \\ \\
{T=4(\Delta _{1} -\Delta _{2} +2\Omega _{2} )\Omega _{2}^{2}\,,} \\ \\
{A=S-(2\Delta _{2} -\Delta _{1} +2\Omega _{1} -4\Omega _{2} )^{2} /3\,,} \\ \\
{B=2(2\Delta _{2} -\Delta _{1} +2\Omega _{1} -4\Omega _{2} )^{3} /27-} \\ {(2\Delta _{2} -\Delta _{1} +2\Omega _{1} -4\Omega _{2} )S/3+T\,, }\\ \\
{D=(A/3)^{3} +(B/2)^{2}\,,} \\ \\ 
{U=(-B/2+\sqrt{D} )^{1/3}\,,} \\ \\
{U^{*} =-(B/2+\sqrt{D} )^{1/3}\,. }
\end{array}
\end{equation*} 

Finally, the three roots are

\begin{equation*}
\begin{array}{l} 
{\mu _{1} =U+U^{*} -(2\Delta _{2} -\Delta _{1} +2\Omega _{1} -4\Omega _{2} )/3\,,} \\ \\
{\mu _{2} =-(U+U^{*} )/2+i\sqrt{3} (U-U^{*})/2\,, }\\ \\
{\mu _{3} =-(U+U^{*} )/2-i\sqrt{3} (U-U^{*})/2\,. }
\end{array} 
\end{equation*}

Then we seek $\alpha _{3} $ in the form

\[\alpha _{3} =c_{1} \; {\rm e}^{i\mu _{1} t} +c_{2} \; {\rm e}^{i\mu _{2} t} +c_{3} \; {\rm e}^{i\mu _{3} t} \] 

\noindent with the initial conditions $\alpha _{3} (0)=0;\; \dot{a}_{3} (0)=0;\; \ddot{a}_{3} (0)=-2\Omega _{1} \Omega _{2} $. These conditions give us the final exact analytical solution for the coefficients,

\[c_{1} =\frac{2\Omega _{1} \Omega _{2} }{(\mu _{2} -\mu _{1} )(\mu _{1} -\mu _{3} )}\,, \] 

\[c_{2} =\frac{2\Omega _{1} \Omega _{2} }{(\mu _{2} -\mu _{1} )(\mu _{3} -\mu _{2} )}\,, \] 

\[c_{3} =\frac{2\Omega _{1} \Omega _{2} }{(\mu _{1} -\mu _{3} )(\mu _{3} -\mu _{2} )}\,. \] 

These formulas allow us to find $\alpha _{3} $ analytically. We can also find $\alpha _{2} $ using Eq.~\eqref{Eq9}:

\[\alpha _{2} =i\dot{\alpha }_{3} e^{i(\Delta _{2} +2\Omega _{1} -2\Omega _{2} )t} /(2\Omega _{2} )\] 

\noindent and thus find the exact analytical solution to Eqs. \eqref{Eq7}-\eqref{Eq9} for arbitrary interaction energy, detunings, and \textit{t}. Taking into account the six-fold level degeneracy, the three-atom Forster resonance spectrum is then calculated as

\[\rho _{3} =(6|a_{2} |^{2} +6|a_{3} |^{2} )/3\]

This value corresponds to the probability to find one of the three atoms in the final $37S_{1/2} $ state and it is the signal measured in our experiments.

\end{document}